\documentclass[fleqn,12pt]{iopart}

\usepackage{iopams}  
\usepackage{graphicx}

\usepackage{amssymb}

  \expandafter\let\csname equation*\endcsname\relax
  \expandafter\let\csname endequation*\endcsname\relax

\usepackage{amsmath,mathtools}

\usepackage{url}
\usepackage{natbib}
\usepackage{appendix}
\usepackage[margin=1cm]{caption}
\usepackage{fullpage}
\usepackage{hyperref}
\usepackage{color}

\newcommand{\Msun}{\ensuremath{M_{ \odot }}}

\mathchardef\mhyphen="2D
\begin{document}

\title{Estimating the sensitivity of pulsar timing arrays}
\author{C. J. Moore, S. R. Taylor and J. R. Gair}
\address{Institute of Astronomy, Madingley Road, Cambridge, CB3 0HA, United Kingdom}
\ead{\mailto{cjm96@ast.cam.ac.uk},\mailto{staylor@ast.cam.ac.uk},\mailto{jrg23@cam.ac.uk}}

\begin{abstract}
The sensitivity curve of a canonical pulsar timing array is calculated for two types of source: a monochromatic wave and a stochastic background. These calculations are performed in both a Bayesian and frequentist framework, using both analytical and numerical methods. These calculations are used to clarify the interpretation of the sensitivity curves and to illustrate the sometimes overlooked fact that the sensitivity curve depends not only on the properties of the pulse time-of-arrival data set but also on the properties of the source being observed. The Bayesian and frequentist frameworks were found to give consistent results and the analytic and numerical calculations were also found to be in good agreement.
\end{abstract}

\section{Introduction}
There is a current global effort under way to detect very low frequency ($f\approx \textrm{yr}^{-1}$) gravitational waves (GWs) via the precision timing of a network of galactic millisecond pulsars. These detection efforts exploit the exquisite rotational stability of millisecond pulsars to track any deviations of the pulse time-of-arrivals (TOAs) from deterministic timing-models. A GW propagating between the Earth and a pulsar will induce a perturbation to the space-time metric along the Earth-pulsar line-of-sight, leading to a change in the proper separation, and consequently a shift in the perceived pulsar rotational frequency \citep{burke-1975,sazhin-1978,detweiler-1979,estabrook-1975}. Subtracting a deterministic timing-model (which describes the pulsar's spin, spin-down rate, etc.) from the TOAs gives a set of {\it timing-residuals}, which encode all unmodeled phenomena, whether they are noise processes or GWs. Utilising a network (or ``array'') of these pulsars allows us to cross-correlate the data-streams and leverage the fact that GWs will be influencing all pulsars \citep{foster-backer-1990,hellings-downs-1983}, whilst intrinsic pulsar noise processes will not. There are three separate pulsar timing array (PTA) efforts underway: the European Pulsar Timing Array (EPTA)\footnote{\url{http://www.epta.eu.org/}} \citep{eptareview2013}, the Parkes Pulsar Timing Array (PPTA)\footnote{\url{http://www.atnf.csiro.au/research/pulsar/ppta/}} \citep{parkesreview2013} and the North American Nanohertz Observatory for Gravitational waves (NANOGrav)\footnote{\url{http://nanograv.org/}} \citep{nanogravreview2013}. There are also ongoing efforts to combine the techniques and data from all three PTAs within the umbrella consortium of the International Pulsar Timing Array (IPTA)\footnote{\url{http://www.ipta4gw.org/}} \citep{iptareview2013}.

The main source of GWs in this frequency band is thought to be a population of supermassive black hole (SMBH) binary systems merging together, with typical masses $\sim 10^8-10^{10} M_{\odot}$ and redshifts $z\lesssim 2$, and in the early, adiabatic inspiral regime of their coalescence \citep{rajaromani1995,jaffe-backer-2003,wyithe-loeb-2003}. It is now well-established that SMBHs are widespread in the nuclei of nearby galaxies \citep[e.g.,][]{ferrarese2005}, with observational relationships such as the famous $M\mhyphen\sigma$ relation indicating symbiotic evolution of the black hole and galactic host \citep[e.g.,][]{ferrarese2000,magorrian1998,marconi2003}. SMBH mergers are expected to be ubiquitous within the currently accepted picture of hierarchical structure formation \citep{whiterees1978,kauffmann2000}, where massive galaxies form via continued accretion from cosmic web filaments, or from galactic mergers. Depending on the distribution of sources in frequency and amplitude these binaries could either be individually resolvable in GWs or overlap to form an unresolved stochastic GW background \citep[e.g.,][]{sesana-vecchio-colacino-2008,sesana-vecchio-volonteri-2009,sesana-vecchio-2010}. Other PTA band sources may include a background from the decay of cosmic-string networks \citep{vilenkin-1981a,vilenkin-1981b,damour-vilenkin-2005,olmez-2010}, or a primordial GW background \citep{grishchuk-1976,grishchuk-2005}, however these are likely to be at a lower amplitude than near-future PTAs will be able to detect.

It is common practice to show the sensitivity of a GW detector by plotting the minimum detectable characteristic-strain as a function of frequency; such plots are known as {\it sensitivity curves} (see Figs.\ \ref{fig:mono_comparison} and \ref{fig:stoch_comparison}). It is commonly assumed, often implicitly, that the sensitivity curve is a function only of properties of the detector. In the case of PTAs the properties of the detector which affect the sensitivity include the error in the timing residuals, cadence, number of pulsars, total duration of observations, etc. If these were the only quantities that affected the sensitivity, then (given a particular PTA) then the sensitivity curve could be calculated once and for all. For any potential source the prospects for detection could be determined by comparing the amplitude of the source and the value of the sensitivity at a given frequency. Such a sensitivity curve {\it cannot} be constructed. Any particular sensitivity curve is not simply a function of the array characteristics, but also the properties of the source. Here, we illustrate this often overlooked subtlety by explicitly calculating sensitivity curves for two different sources: a monochromatic wave and a power-law stochastic background.

For both of these sources the sensitivity curve was calculated in three ways: (1) using a simple analytic treatment based on a frequentist definition of detection; (2) using a simple analytic treatment based on a Bayesian definition of detection; (3) employing a full numerical analysis on mock datasets using a Bayesian data-analysis pipeline. The two main aims of this paper are to illustrate the differences between the sensitivity curves for different sources and to demonstrate the consistency of the Bayesian and frequentist approaches. However as a by-product we derive simple analytic formulae for the sensitivity curves, and the scaling of the signal-to-noise ratio (SNR) and Bayesian-evidence with the characteristics of the PTA. It is hoped that these simple analytic sensitivity curves, when combined with predicted source distributions from different scenarios, will be useful for evaluating prospects for detection.

In Section\ \ref{sec:PTA} the response of a PTA to incident GWs is briefly reviewed, followed by calculations in Sections\ \ref{sec:mono} and \ref{sec:stoch} of the sensitivity of a PTA to a monochromatic source and a stochastic background.

\section{The response of a PTA}\label{sec:PTA}
The Earth and all pulsars reside in the metric perturbation field
\begin{equation} h_{ab}(t,\vec{x})=\sum_{A=+,\times}\int\textrm{d}f\;\iint_{S_{2}}\textrm{d}\hat{\Omega}\; \tilde{h}_{A}(f,\hat{\Omega})e_{ab}^{A}(\hat{\Omega})\exp\left( 2\pi i f (t-\hat{\Omega}\cdot\vec{x}) \right) \; ,\end{equation}
where $\vec{x}$ is the spatial position and $\tilde{h}_{A}(f,\hat{\Omega})$ is the Fourier amplitude of the GW of frequency $f$ propagating in direction $\hat{\Omega}$. Taking the Earth to be at the origin of the coordinate system, and for a pulsar $x$ at position $\bold{p}_{x}$, the measured quantity is the timing-residual, corresponding to GW-induced deviations of the TOAs from those computed by a deterministic timing-model, and given by
\begin{equation}\label{eq:relateresidualtoredshifts} R_{x}(t,\hat{\Omega})=\int_{0}^{t}\textrm{d}t'\;z_{x}(t',\hat{\Omega}) \; , \quad \textrm{where} \; x\in \left[1,N_{p}\right]\, ,\end{equation}
where $z_{x}(t)$ is the redshift of the rate of arrival of signals from pulsar $x$ induced by the gravitational waves and $N_{p}$ is the number of pulsars in the PTA. This depends on the metric perturbation at the Earth and at the pulsar \citep{anholm-2009,brook-flanagan-2011}
\begin{equation} \label{eq:pultermandEterm}z_{x}(t,\hat{\Omega})=\frac{1}{2}\frac{\hat{p}_{x}^{j}\hat{p}_{x}^{i}} {1+\hat{\Omega}\cdot\hat{p}_{x}}\left(h_{ij}(t_{p,x},\hat{\Omega} )-h_{ij}(t_{e},\hat{\Omega} )\right) \, ,\end{equation}
where $t_{e}=t$, $t_{p,x}=t-L_x(1+\hat{\Omega}\cdot\hat{p}_{x})$, and $L_x$ is the distance to the pulsar. In the following the pulsar-term (the first term in Eq.\ \ref{eq:pultermandEterm}) is ignored as self-noise which averages to zero when calculating correlations between pulsar residuals. Within this approximation the residual depends solely on the local metric perturbation at the Earth. Throughout this paper the timing residuals will be treated as the measured signal, it is also possible to work with the redshifts related to the timing residuals by Eq.\ \ref{eq:relateresidualtoredshifts}.

When producing a particular sensitivity curve it is necessary to assume particular values for various PTA quantities. For the analytic calculations in this paper a canonical PTA consisting of 36 pulsars distributed randomly on the sky, timed fortnightly to a precision of $100\,\textrm{ns}$ over a total baseline of $5\,\textrm{years}$ was assumed. This is roughly equivalent to mock dataset \textsc{Open1} in the recent IPTA data challenge\footnote{\url{http://www.ipta4gw.org/?page_id=89}}, the characteristics of which were used to produce the numerical sensitivity curves. It should be noted that this mock dataset is much more sensitive than that of any current PTA. It is straighforward to generalise the analysis to more complicated situations where each pulsar has a different cadence, timing precission, length of observation, etcetera.

\section{The monochromatic source: e.g. a non-evolving binary}\label{sec:mono}

\subsection{Frequentist detection}\label{sec:mono_freq}
The frequentist method involves defining a detection statistic ${\cal{S}}$. The SNR of this statistic is defined as the expectation in the presence of a signal divided by the root mean square (rms) value in the absence of a signal. A detection is claimed if the SNR in a particular realisation of the experiment exceeds a predetermined threshold value $\varrho_{\textrm{th}}$. Here a threshold of $\varrho_{\textrm{th}}=3$ was used. 

The noise in the timing-residuals is assumed to be white, Gaussian, and uncorrelated between each pulsar. 
Let $s_{x}(t)$ be the smooth function from which the discretely sampled timing risiduals in pulsar $x$ are drawn. The real data from a PTA contain noise and a signal
\begin{equation} {\bf{s}}(t)={\bf{n}}(t)+{\bf{h}}(t) \; , \quad \textrm{where }\,{\bf{s}}(t)^{\textrm{T}}=\left(s_{1}(t),s_{2}(t), \,\ldots ,s_{N_{p}}(t)\right)\, .\end{equation}
The noise satisfies $\left<\tilde{n}_{x}(f)\tilde{n}_{y}^{*}(f')\right>=(1/2)\delta (f-f')\delta_{xy}S_{n,x}$, where $S_{n,x}=2\sigma_{x}^{2}\delta t_{x}$, $1/\delta t_{x}$ and $\sigma_{x}$ are the cadence and the rms error in the timing-residuals in pulsar $x$. Cross correlating the residuals with a symmetric filter matrix ${\bf{K}}(t)$ defines a statistic and associated SNR
\begin{eqnarray}{\cal{S}}&=\int\textrm{d}t\,\int\textrm{d}t'\,{\bf{s}}(t)^{\textrm{T}}{\bf{K}}^{\dagger}(t-t'){\bf{s}}(t')=\int\textrm{d}f\;\tilde{{\bf{s}}}(f)^{\textrm{T}}\tilde{{\bf{K}}}^{\dagger}(f)\tilde{{\bf{s}}}(f)  \;, \\
 \varrho^{2} &= \frac{\mu^{2}}{\sigma^{2}}=\frac{\left<{\cal{S}}\right>_{s=h+n}^{2}}{\left<{\cal{S}}^{2}\right>_{s=n}-\left<{\cal{S}}\right>_{s=n}^{2}}\; . \label{e:monosnr}\end{eqnarray}
Using the identical Gaussian properties of the noise in each pulsar it is straightforward to show that the expectation value of ${\cal{S}}$ in the presence of a signal and the variance of ${\cal{S}}$ in the absence of a signal are given respectively by
\begin{eqnarray} \mu &= \int\textrm{d}f\, \left[ \tilde{{\bf{h}}}(f)^{\textrm{T}}\tilde{{\bf{K}}}^{\dagger}(f)\tilde{{\bf{h}}}(f) + \frac{T}{2}S_{n}\textrm{Tr}\left(\tilde{{\bf{K}}}^{\dagger}\right) \right] \, , \label{eq:mu}\\
\sigma^{2}&=\int\textrm{d}f\,\left[ \frac{T}{4}S_{n}^{2}\textrm{Tr}\left(\tilde{{\bf{K}}}(f)\tilde{{\bf{K}}}(f)^{\dagger}\right) \right] -\left(\int\textrm{d}f\,\frac{T}{2}S_{n}\textrm{Tr}\left(\tilde{{\bf{K}}}^{\dagger}\right)\right)^{2} \, . \label{eq:sigma2}\end{eqnarray}
The contributions to ${\cal{S}}$ from pulsar auto-correlations are neglected, this is achieved by setting the diagonal elements of $\tilde{{\bf{K}}}(f)$ to zero, so that the $\textrm{Tr}(\tilde{{\bf{K}}}^{\dagger})$ in Eqs.\ \ref{eq:mu} and \ref{eq:sigma2} vanish. The SNR, Eq.~(\ref{e:monosnr}), is then given by the following inner product 
\begin{eqnarray}\label{eq:whatisop} \varrho^{2} = \frac{4}{T}\frac{\left(\frac{   \tilde{ {\bf{h}} }(f)\tilde{ {\bf{h}} }(f)^{\textrm{T}}}{S_{n}^{2}}|\tilde{{\bf{K}}}(f)\right)^{2}}{\left(\tilde{{\bf{K}}}(f)|\tilde{{\bf{K}}}(f)\right)} \, , \quad \textrm{where}\;\left(\tilde{{\bf{A}}}|{\tilde{\bf{B}}}\right)=\int\textrm{d}f\,\textrm{Tr}\left(\tilde{{\bf{A}}}\tilde{{\bf{B}}}^{\dagger}\right)S_{n}^{2} \, . \end{eqnarray}
The optimal filter is the matrix ${\bf{K}}(t)$ which maximises the SNR in Eq.\ (\ref{eq:whatisop}). It follows from the Cauchy-Schwarz inequality that the optimal filter (up to an arbitrary factor) and its corresponding SNR are given by
\begin{eqnarray}
&\tilde{{\bf{K}}}(f)=\left. \frac{\tilde{{\bf{h}}}(f)\tilde{{\bf{h}}}(f)^{\textrm{T}}}{S_{n}^{2}}\right|_{\textrm{diag}\rightarrow 0}\; , \quad  \textrm{where}\;\varrho^{2}=\sum_{y}\sum_{x>y}\frac{8}{T}\int\textrm{d}f\;\frac{\left|\tilde{h}_{x}(f)\right|^{2}\left|\tilde{h}_{y}(f)\right|^{2}}{S_{n}^{2}} \, \label{eq:SNRmono}.
\end{eqnarray}
In order to evaluate the optimal, or ``matched'',  filter $\tilde{{\bf{K}}}(f)$, the waveform $\tilde{{\bf{h}}}(f)$ must be known to sufficient accuracy. A monochromatic source is a simple example of a waveform which can be extracted using matched filtering. The expression for $\varrho^{2}$ in Eq.\ \ref{eq:SNRmono} is different from the usual expression for the SNR of a matched filter search, in particular it scales as $\sim \tilde{h}^{4}$ instead of $\sim\tilde{h}^{2}$. This is because our detection statistic in Eq.\ \ref{e:monosnr} cross-correlates the signals from different pulsars. It would be possible to use the standard matched filter statistic, however the cross-correlation statistic has the advantage that it makes the single source search directly comparable to the stochastic background search (see Sec.\ \ref{sec:stoch}), where one is forced to used a cross-correlation statistic because the stoachastic signal in each pulsar is buried in the pulsar red-noise. The two searches have comparable sensitivities.

From Eq.\ (\ref{eq:pultermandEterm}) it may be seen that the measured signal is proportional to the GW amplitude and a geometric factor depending on the sky positions of the pulsars, the sky position of the source and the source orientation. Since there are many pulsars in our PTA they effectively average this distribution over the sky position angles. For simplicity we set the source inclination and polarisation angles to zero, so we are calculating the sky-averaged sensitivity of the PTA to optimally orientated sources. It is straighforward to generalise this treatment to arbitrary source orientations. The sky-averaged value of the geometric factor in Eq.\ (\ref{eq:pultermandEterm}) is
\begin{eqnarray} \label{eq:chiavoversky} \chi &= \int_{\phi=0}^{2\pi} \int_{\theta=0}^{\pi}\,\frac{\sin\theta\textrm{d}\theta\textrm{d}\phi}{4\pi}\, \sqrt{\left(\frac{1}{2}\frac{\hat{p}_{i}\hat{p}_{j}\left(A^{+}H^{+}_{ij}+A^{\times}H^{\times}_{ij}\right)}{1+\hat{\Omega}\cdot\hat{p}}\right)^{2}} = \frac{1}{\sqrt{3}}\, ,\end{eqnarray}
where $\hat{p}=\left(\sin\theta\cos\phi,\sin\theta\sin\phi,\cos\theta\right)$, $e^{x}=(1,0,0)$, $e^{y}=(0,1,0)$, $\hat{\Omega}=\left(0,0,1\right)$, $H^{+}_{ij}=\epsilon^{+}_{ij}\cos 2\psi+\epsilon^{\times}_{ij}\sin 2\psi$, $H^{\times}_{ij}=-\epsilon^{+}_{ij}\sin 2\psi+\epsilon^{\times}_{ij}\cos 2\psi$, $\epsilon^{+}_{ij}=e^{x}_{i}e^{x}_{j}-e^{y}_{i}e^{y}_{j}$, $\epsilon^{\times}_{ij}=e^{x}_{i}e^{y}_{j}+e^{y}_{i}e^{x}_{j}$, $A^{+}=(1+\cos^{2}\iota)/2$, $A^{\times}=\cos\iota$ and $\iota=\psi=0$. Therefore for a monochromatic source of GWs with frequency $f_{0}$ the signal is given approximately by $\tilde{h}_{x}(f)\approx\tilde{h}_{y}(f)\approx(\chi h_{c}/f)\delta (f-f_{0})$. Using Eq.\ (\ref{eq:SNRmono}), together with the finite-time delta function, $\delta_{T}(f)=\sin\left(\pi fT\right)/(\pi f)$, gives
\begin{equation}\label{eq:1} \varrho^{2}=\frac{1}{2}N_{p}\left(N_{p}-1\right)\frac{8\chi^{4}h_{c}^{4}}{T}\int_{1/T}^{1/\delta t}\textrm{d}f\;\frac{\delta^{4}_{T}(f-f_{0})}{f^{4}S_{n}^{2}} \; . \end{equation}
The PTA is sensitive to frequencies in the range $\sim 1/T$ up to the Nyquist frequency. Imposing a threshold for detection, $\varrho=\varrho_{\textrm{th}}$ and rearranging gives $h_{c}$ as a function of $f_{0}$. This is the desired sensitivity curve, and is shown as the red curve in the left panel of Fig.\ \ref{fig:mono_comparison}. The fact that the sensitivity curve tends to a constant value at low frequencies is obviously incorrect. The reason for this is that the loss of sensitivity which arises from fitting a deterministic timing-model to the raw TOAs has not been accounted for. Or, in Bayesian language, Eq.\ (\ref{eq:1}) assumes delta function priors on all of the pulsar timing model parameters.

\subsubsection{Time domain}\label{sec:mono_freq_time}
Some insight into the shape of the sensitivity curve and the loss of sensitivity due to fitting for the pulsar timing-model may be gained by considering the inner product in the time domain. In Eq.\ (\ref{eq:1}), if the power of $4$ were replaced by a power of $2$ the fact that the noise is white would allow us to use Parseval's thoerem to change from a frequency integral to a time integral. By analogy, from Eq.\ (\ref{eq:1}) the SNR may be written approximately as
\begin{equation} \varrho^{2}\approx \frac{1}{2}N_{p}\left(N_{p}-1\right)T\int_{0}^{T}\textrm{d}t\;\frac{\chi^{4}h_{c}^{4}\sin^{4}\left(2\pi ft+\phi\right)}{\sigma^{4}f^{4}\delta t^{2}}  \;.\end{equation}
For our PTA, at frequencies of $\approx 1\,\textrm{yr}^{-1}$, this approximation holds to better than $10\%$. In the high frequency limit ($ft\gg 1$) the integral $\int\sin^{4}\left(2\pi ft+\phi\right)\,\textrm{d}t\approx 3T/8$ and the sensitivity tends to
\begin{equation}\label{eq:high}  h^{\textrm{HIGH}}_{c}(f)\approx \left(\frac{16\varrho_{\textrm{th}}^{2}}{3\chi^{4}N_{p}\left(N_{p}-1\right)}\right)^{1/4}\sigma f \sqrt{\frac{\delta t}{T}} \;.\end{equation}
In the low frequency limit ($ft\ll 1$) the sine may be expanded as a power series
\begin{equation} \sin\left(2\pi ft+\phi\right)\approx \sin (\phi)+2\pi tf\cos(\phi)-2\pi^{2}f^{2}t^{2}\sin (\phi)-\frac{4}{3}\pi^{3}f^{3}t^{3}\cos (\phi )+{\cal{O}}\left(f^{4}t^{4}\right)\; .\end{equation}
The first term in this expansion is a constant offset from zero in the timing residuals; this type of signal is degenerate with the distance to the pulsar and hence will be ``fit-out'' of the data. Similarly the second term is a linear drift in the residuals and is degenerate with the pulsar spin (and also the line of sight component of the peculiar velocity). Finally the third term is degenerate with the pulsar spin-down rate. Higher order derivatives of the spin period do not need to be fit for independently as they are determined uniquely by the period and its first time derivative. Therefore at low frequencies the leading contribution, at ${\cal{O}}\left(f^{3}t^{3}\right)$, gives
\begin{equation}\label{eq:low}  h^{\textrm{LOW}}_{c}(f)\approx \frac{3\sqrt{\varrho_{\textrm{th}}}}{2^{7/4}\chi\pi^{3}}\left(\frac{13}{N_{p}(N_{p}-1)}\right)^{1/4}\sigma f^{-2}\sqrt{\frac{\delta t}{T}}T^{-3}\sec\phi \;.\end{equation}
The total sensitivity may be approximated by $h_{c}^{\textrm{LOW}}+h_{c}^{\textrm{HIGH}}$, i.e. as a two part power-law in $f$. This is shown as the blue curve in the left panel of Fig.\ \ref{fig:mono_comparison}, where the value of $\phi$ has been chosen to make $h_{c}^{\textrm{LOW}}=h_{c}^{\textrm{HIGH}}$ at a frequency of $2/T$. Despite the apparent simplicity of this two part power law model it shows excellent agreement with the following Bayesian treatment.

\subsection{Bayesian detection}\label{sec:mono_baye}
One advantage of the Bayesian approach is that it provides a well motivated method for accounting for the loss of sensitivity due to fitting for the pulsar timing-model, as opposed to the rather \emph{ad hoc} subtraction of a few terms from a Taylor series performed above.

In the Bayesian approach two competing hypotheses are considered: the noise and signal hypotheses. The noise hypothesis, ${\cal{H}}_{n}$, is that the data contain only contributions from noise and the timing-model while the signal hypothesis, ${\cal{H}}_{h}$, is that the data contain noise, timing-model and a signal. For each hypothesis, $i \in \left\{h,n\right\}$, the evidence may be calculated as 
\begin{equation} {\cal{O}}_{i}({\bf{s}})=\int\textrm{d}\vec{\lambda}\;{\cal{L}}_{i}({\bf{s}},\vec{\lambda}_{i})P_{i}(\vec{\lambda}_{i}) \; ,\end{equation}
where $\vec{\lambda}_{i}$ is the vector of free parameters, ${\cal{L}}_{i}$ is the likelihood function, ${\bf{s}}$ is the measured data and $P_{i}$ is the prior function, for hypothesis ${\cal{H}}_{i}$. From here on the dependence on the data is suppressed in our notation for compactness. The Bayes factor is then defined as the ratio of these evidences, and a detection is claimed if this exceeds some pre-determined threshold, ${\cal{B}}\equiv {\cal{O}}_{h}/{\cal{O}}_{n}>{\cal{B}}_{\textrm{th}}$. A value of ${\cal{B}}_{\textrm{th}}=1000$ was used, this was choosen to give roughly the same false alarm rate at the value of $\varrho_{\textrm{th}}=3$ used in Sec. \ref{sec:mono_freq}.

In this section the physical signal, $h_{x}(t)$, is assumed to be that of a monochromatic source, and a quadratic timing model, $m_{x}(\vec{\Theta}_{x},t)$, for each pulsar is assumed, with pulsar parameters $\vec{\Theta}_{x}$. In reality the timing model is more complex than a simple quadratic as it has to account for several effects, such as the pulsar's position, dispersion in the interstellar medium, peculiar motion, and orbital motion if the pulsar happens to be in a binary system \citep[see][and references therein]{tempo2-1,tempo2-2}. However, a simple quadratic model serves here to illustrate the loss of sensitivity due to fitting for the distance to the pulsar, the pulsar spin and the pulsar spindown. The quadratic model also has the nice property that the pulsar timing-model parameters can be marginalised over analytically,
\begin{eqnarray} 
&{\bf{s}}=( \overbrace{\underbrace{s_{1}(\delta t),s_{1}(2\delta t),\ldots,s_{1}(T)}_{T/\delta t},  \ldots,s_{N_{p}}(\delta t),s_{N_{p}}(2\delta t),\ldots,s_{N_{p}}(T)}^{N_{p}T/\delta t} ) \, , \nonumber\\
&h_{x}(\vec{\Psi},t)=\frac{\chi h_{c}}{f}\sin\left(2\pi f t +\phi \right)\, , \quad \textrm{with source parameters}\;\vec{\Psi}^{\textrm{T}}=\left\{h_{c},f,\phi\right\},\label{eq:bigeqn}\\
&m_{x}(\vec{\Theta}_{x},t)= \vec{\Theta}^{\textrm{T}}_{x}\cdot\vec{N}\, , \quad\quad\quad\quad  \textrm{with }\vec{N}^{\textrm{T}}=\left\{1,t,t^{2}\right\}\;\textrm{and}\;\vec{\Theta}^{\textrm{T}}_{x}=\left\{\alpha_{x},\beta_{x},\gamma_{x}\right\}. \nonumber\end{eqnarray}
where $\alpha_x$ is a constant phase offset, $\beta_x$ is proportional to the pulsar's rotational frequency or peculiar velocity, and $\gamma_x$ is proportional to its spindown rate or acceleration. Since the noise is Gaussian the log-likelihood for the noise hypothesis is given by 
\begin{equation}\label{eq:logLn} \log {\cal{L}}_{n}(\vec{\Theta})=\log A-\frac{1}{2}\left({\bf{s}}-{\bf{m}}(\vec{\Theta})\right)^{\textrm{T}}{\bf{\Sigma}}_{n}^{-1}\left({\bf{s}}-{\bf{m}}(\vec{\Theta})\right)\;, \end{equation}
where the covariance matrix is simply the scaled identity, ${\bf{\Sigma}}_{n}=\sigma^{2}{\bf{I}}_{N_{p}T/\delta t}$, and $A$ is a constant, absorbing determinant factors. Similarly the log-likelihood for the signal hypothesis is given by
\begin{eqnarray}\label{eq:logLh}  \log {\cal{L}}_{h}(\vec{\Theta},\vec{\Psi})=\log A-\frac{1}{2}\left({\bf{s}}-{\bf{m}}(\vec{\Theta})-{\bf{h}}(\vec{\Psi})\right)^{\textrm{T}}{\bf{\Sigma}}_{n}^{-1}\left({\bf{s}}-{\bf{m}}(\vec{\Theta})-{\bf{h}}(\vec{\Psi})\right)\;.\end{eqnarray}

In both the noise and signal hypotheses uniform priors on the timing-model parameters were assumed. In the signal hypothesis case there are also the priors on the source parameters to consider. Since we have adopted a very stringent detection threshold (Bayes factor of $1000$) it is reasonable to expect the posterior to be strongly peaked at the true values independent of any (reasonable) prior used. Of course the data must be used to find the position of this peak and this process will exhaust a certain amount of information in the data reducing the final evidence value. However, this reduction may be neglected in the limit of large final evidence. Numerically this approximation is equivalent to taking a delta-function prior on the source parameters positioned at the correct values, $P(\vec{\Psi},\vec{\Theta}_{x})\propto\delta^{(3)}(\vec{\Psi}-\vec{\Psi}')$, however it should be stressed that this is an analytic trick used to implement the approximation described and in practice the data will still be used to find the maximum in the posterior. (It may be the case that a very localised prior is used on the source parameters if a clear electromagnetic counterpart has been identified.)

From Eq.\ \ref{eq:bigeqn} it can be seen that the timing model is linear in the pulsar parameters, so the timing model may be expressed by a $(N_{p}T/\delta t)\times(3N_{p})$ matrix, ${\bf{M}}$, known as the design matrix. This non-square matrix admits the usual unique singular value decomposition into the $(N_{p}T/\delta t)\times(N_{p}T/\delta t)$ matrix ${\bf{U}}$, the $(N_{p}T/\delta t)\times(3N_{p})$ matrix ${\bf{S}}$ and the $(3N_{p})\times(3N_{p})$ matrix ${\bf{V}}$. The matrix ${\bf{U}}$ may be further uniquely decomposed into ${\bf{F}}$ and ${\bf{G}}$ where ${\bf{G}}$ is an $(N_{p}T/\delta t)\times(N_{p}T/\delta t - 3N_{p})$ matrix.
\begin{equation} {\bf{m}}={\bf{M}}\vec{\Theta}\; , \quad   {\bf{M}}={\bf{U}}{\bf{S}}{\bf{V}}^{\dagger}\; , \quad  {\bf{U}}=\left({\bf{F}},{\bf{G}}\right)\, . \end{equation}

The evidence integral for the noise hypothesis is a multivariate Gaussian in the pulsar timing-model parameters, $\vec{\Theta}_{x}$. This may be evaluated analytically \citep{van-haasteren-levin-2012}, and can be viewed as a projection of the data into the left null space of the design matrix.
\begin{equation} \label{eq:On} {\cal{O}}_{n}=\int\textrm{d}\vec{\Theta}\; {\cal{L}}_{n}(\vec{\Theta}) =\frac{\exp\left(-\frac{1}{2}{\bf{s}}^{\textrm{T}}{\bf{G}}\left({\bf{G}}^{\textrm{T}}{\bf{\Sigma}}_{n}{\bf{G}}\right)^{-1}{\bf{G}}^{\textrm{T}}{\bf{s}}\right)}{\sqrt{(2\pi)^{n-m}\textrm{det}\left({\bf{G}}^{T}{\bf{\Sigma}}_{n}{\bf{G}}\right)}}\end{equation}
The evidence for the signal hypothesis may be similarly calculated by evaluating the following integral,
\begin{equation} {\cal{O}}_{h}= \int\textrm{d}\vec{\Psi}\;\delta(\vec{\Psi}-\vec{\Psi}')\int\textrm{d}\vec{\Theta}\; {\cal{L}}_{h}(\vec{\Theta}_{x},\vec{\Psi}).\end{equation}
However, as can be seen by comparing Eqs.\ (\ref{eq:logLn}) and (\ref{eq:logLh}), this is identical to the result in Eq.\ (\ref{eq:On}) with the transformation ${\bf{s}}\rightarrow {\bf{s}}-{\bf{h}}(\vec{\Psi}')\equiv {\bf{s}}-{\bf{h}}'$. The Bayes factor, defined as the ratio of the two evidences times the prior odds ratio, is then given by the following (where the prior odds has been set to unity),
\begin{equation}\label{eq:temp} {\cal{B}}\equiv\frac{{\cal{O}}_{h}}{{\cal{O}}_{n}}= \exp\left( -\frac{1}{2}{\bf{h}}'^{\textrm{T}} {\bf{G}}\left({\bf{G}}^{\textrm{T}}{\bf{\Sigma}}_{n}{\bf{G}}\right)^{-1}{\bf{G}}^{\textrm{T}}{\bf{h}}' +\frac{1}{2}{\bf{s}}^{\textrm{T}} {\bf{G}}\left({\bf{G}}^{\textrm{T}}{\bf{\Sigma}}_{n}{\bf{G}}\right)^{-1}{\bf{G}}^{\textrm{T}}{\bf{s}} \right)\, .  \end{equation}
The measured data is given by ${\bf{s}}={\bf{h}}'+{\bf{m}}+{\bf{n}}$. We also have by definition of the projection matrix ${\bf{G}}^{\textrm{T}}{\bf{m}}=0$ . 
Averaging the Bayes factor over many realisations of Gaussian noise with gives the expectation value of the Bayes factor as $\overline{\cal{B}}$.
\begin{eqnarray} &P({\bf{n}})\textrm{d}{\bf{n}}=\frac{\exp\left(-\frac{1}{2}{\bf{n}}^{\textrm{T}}{\bf{\Sigma}}_{n}^{-1}{\bf{n}}\right)}{\sqrt{(2\pi)^{N_{p}T/\delta t}\textrm{det}\left({\bf{\Sigma}}_{n}\right)}}\,\textrm{d}{\bf{n}}\, , \\
&\label{eq:final} \overline{{\cal{B}}}=\int\textrm{d}{\bf{n}}\;P({\bf{n}}){\cal{B}}=\exp\left( \left({\bf{G}}^{\textrm{T}}{\bf{h}}'\right)^{\textrm{T}} \left({\bf{G}}^{\textrm{T}}{\bf{\Sigma}}_{n}{\bf{G}}\right)^{-1} {\bf{G}}^{\textrm{T}}{\bf{h}}' \right) \; .\end{eqnarray}

Hence the expected value of the Bayes factor is given by inner product of the signal, projected orthogonal to the quadratic timing model, with itself. This inner product may be written as integral in the time domain where the physical signal, $h(t)$, is replaced with the projected signal, $(Gh)(t)$. The projection is accomplished by explicitly choosing a basis of three orthogonal function which span the space of the quadratic timing model.
\begin{eqnarray}\label{eq:absfinal} &\overline{{\cal{B}}}=\exp\left(N_{p}\int_{0}^{T}\textrm{d}t \, \frac{(Gh)(t)^{2}}{2\sigma^{2}\delta t}\right)\quad \textrm{where,} \\
& (Gh)(t)=h(t)-f_{1}(t)\frac{\int_{0}^{T}\textrm{d}\tau\,f_{1}(\tau)h(\tau)}{\int_{0}^{T}\textrm{d}\tau\,f_{1}(\tau)^{2}}-f_{2}(t)\frac{\int_{0}^{T}\textrm{d}\tau\,f_{2}(\tau) h(\tau)}{\int_{0}^{T}\textrm{d}\tau\,f_{2}(\tau)^{2}}-f_{3}(\tau)\frac{\int_{0}^{T}\textrm{d}\tau\,f_{3}(t)h(\tau)}{\int_{0}^{T}\textrm{d}\tau\,f_{3}(\tau)} \nonumber \\ 
&f_{1}(\tau)=\frac{\tau^{2}}{T}-\tau+\frac{T}{6},\quad f_{2}(\tau)=\tau-\frac{T}{2} \quad\textrm{and}\quad f_{1}(\tau)=T \nonumber .\end{eqnarray}
Setting $\overline{{\cal{B}}}={\cal{B}}_{\textrm{th}}$ in Eq.\ \ref{eq:absfinal} and rearranging gives an expression for $h_{c}(f,\phi)$, which defines the sensitivity curve (this expression is somewhat lengthy, see \ref{app}). The final three terms in the expression for $(Gh)(t)$ arose from marginalising over the timing-model, neglecting these terms gives another expression, $H_{c}(f,\phi)$, which corresponds to the sensitivity without the loss due to a timing-model fit. Both of these are plotted in the right panel of Fig.\ \ref{fig:mono_comparison} for different values of $\phi$. This again illustrates the loss in sensitivity due to the requirement that we fit for free parameters in the timing-model. The black curves in Fig.\ \ref{fig:mono_comparison} are the phase averaged sensitivities.

\subsection{Numerical calculations}\label{sec:mono_num}
The shape of the strain sensitivity curve is now reconstructed with numerical simulations of {\it Earth-term only} monochromatic signal injections. For our canonical PTA, we adopt the $36$ pulsar network of the first IPTA data challenge, where pulsars are timed fortnightly to $100$ ns precision over a total baseline of $5$ years. Injections are performed using the \textsc{PALSimulation} code, which is part of the \textsc{PAL} package\footnote{\url{https://github.com/jellis18/PAL}} being developed as a unifying suite of tools for pulsar timing analysis. These injections provide a set of simulated timing files, which along with associated pulsar parameter files, can be processed with the \textsc{Tempo2} pulsar-timing package \citep{tempo2-1,tempo2-2,tempo2-3}. The output from the \textsc{Tempo2} timing-model fit is a set of timing-residuals, and the design-matrix which describes the contribution of the deterministic timing-model parameters to each TOA observation. The likelihood model was analytically marginalised over uniform-prior timing-model parameters by projecting all quantities into the left null-space of the design matrix, equivalent to a linear operation on the timing-residuals and noise matrices \citep{van-haasteren-levin-2012}.

For a grid of PTA-band frequencies and GW-source distances, a systematic injection and recovery of varying SNR Earth-term only signals was performed. Searches were over the $7$-dimensional parameter space of $\{\zeta,f,\theta,\phi,\iota,\psi,\phi_0\}$, where $\zeta=\mathcal{M}^{5/3}/D_L$ is a dimensionless strain-amplitude defined in terms of the binary chirp mass, $\mathcal{M}$, and luminosity distance, $D_L$; $f$ is the GW frequency; $(\theta,\phi)$ denote the sky-location of the source in spherical-polar coordinates; $\iota$ is the orbital inclination angle; $\psi$ is the GW polarisation angle; and $\phi_0$ is an initial orbital phase parameter. The angles $\phi_{0}$, $\psi$ and $\iota$ for the source were set to be zero; the sky position angles were set as 
%$\phi=1^{\textrm{rad}}$ and $\theta=\pi/2-0.5^{\textrm{rad}}$. 
$\phi=1$ and $\theta=\pi/2-0.5$. 
For the sky positions of the 36 pulsars in the PTA the root mean square value of the geometric factor in the integrand of Eq.\ \ref{eq:pultermandEterm} is $\left<\chi\right>=0.51$. This is in reasonable agreement with the expected value calculated in Eq.\ \ref{eq:chiavoversky} of $1/\sqrt{3}\approx 0.58$. The chirp mass was set as $\mathcal{M}=10^{7}\Msun$; and the luminosity distance was varried between $10^{-5}\,\textrm{Mpc}$ and $10\,\textrm{Mpc}$. This choice of source parameters ensures that the ``chirping'' timescale of the binary due to orbital shrinkage by GW-emission is much longer than the baseline of 5 yrs, whilst the range of distances scales the SNR of the injection from the regime of being completely undetectable to easily detectable.

\begin{figure}[h!]
 \centering
 \includegraphics[trim=0cm 0cm 0cm 0cm, width=0.9\textwidth]{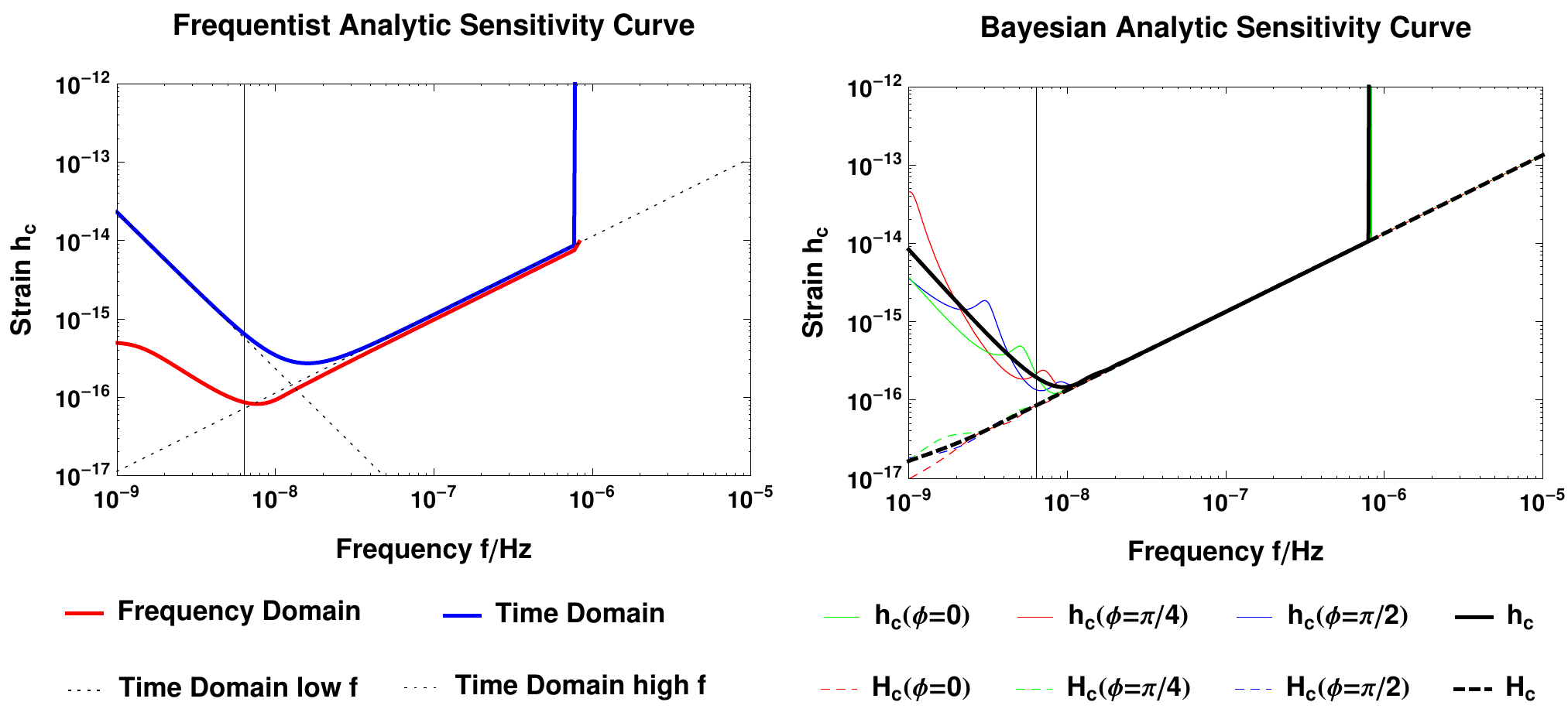}
 \caption{Sensitivity curves for the PTA discussed in the text to a monochromatic source. The left panel shows the prediction of the frequentist formula in Sec.\ \ref{sec:mono_freq}, the right panel shows the prediction of the Bayesian formula in Sec.\ \ref{sec:mono_baye}.}
 \label{fig:mono_comparison}
\end{figure}

\begin{figure}[h!]
 \centering
 \includegraphics[trim=0cm 0cm 0cm 0cm, width=0.9\textwidth]{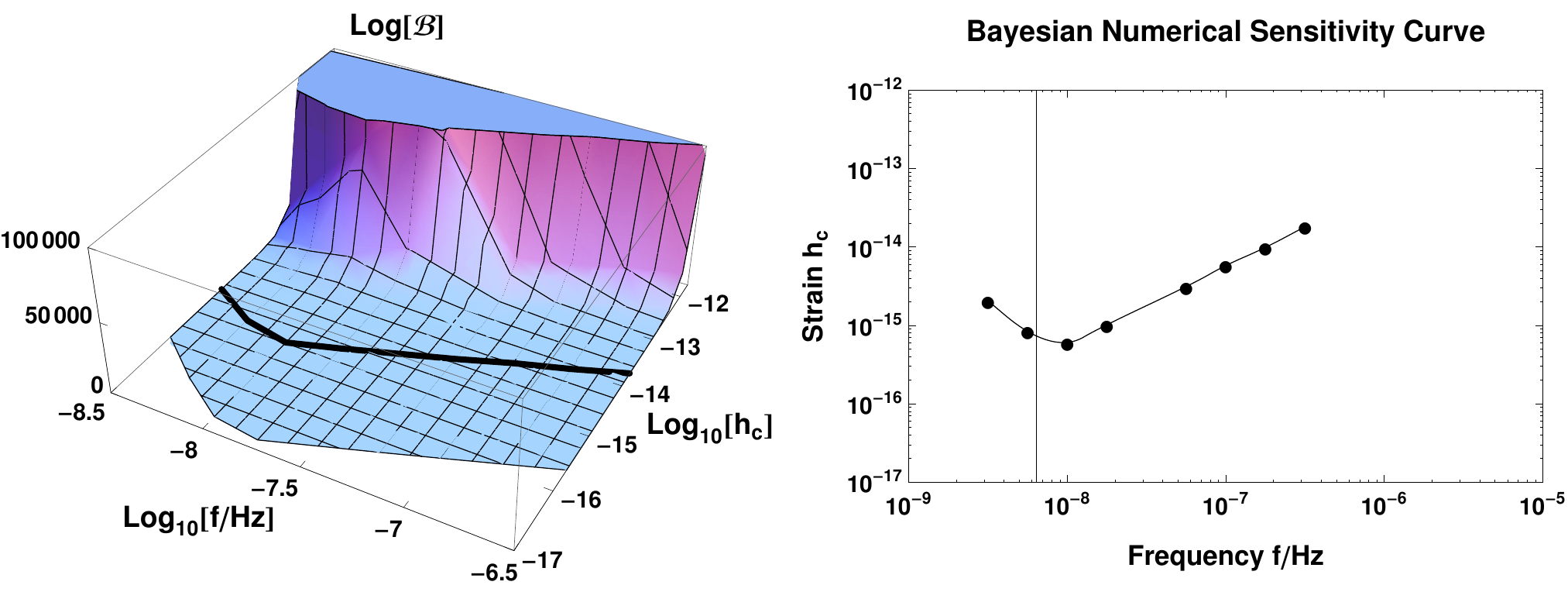}
 \caption{The left panel shows a plot of $\log ({\cal{B}})$ against amplitude and frequency. The black line is the contour ${\cal{B}}={\cal{B}}_{\textrm{th}}$. The black curve is identical to that plotted in the right panel which shows the numerically calculated sensitivity curve in Sec.\  \ref{sec:mono_num}.}
 \label{fig:mono_surface}
\end{figure}

Parameter estimation and evidence recovery are performed using the Bayesian inference package \textsc{MultiNest} \citep{feroz2008,feroz2009,importanceMultiNest2013}. The collection of recovered Bayesian evidence values were interpolated at each injected frequency to determine the characteristic strain-amplitude at which we exceed the pre-determined detection threshold. The Bayes factor surface is shown in the left panel of Fig.\ \ref{fig:mono_surface}, along with our numerically deduced Bayesian sensitivity curve in the right panel of Fig.\ \ref{fig:mono_surface}. Comfortable qualitative agreement can be seen with the results of the simple frequentist/Bayesian analytic techniques shown in Fig.\ \ref{fig:mono_comparison}.

\section{The stochastic background: e.g. a superposition of binaries}\label{sec:stoch}

\subsection{Frequentist detection}\label{sec:stoch_freq}
When searching for a stochastic background it is not possible to use the above statistic as it requires the matched signal templates. The best that can be done is to predict the statistical properties of the signal. A power-law Gaussian stochastic background is characterised by an amplitude and a slope, $\vec{\Psi}=\left\{A,\alpha\right\}$, where the second moment of the Fourier amplitudes completely defines the statistical properties of the timing residuals
\begin{equation}\label{eq:OverlapReductionFunction} \left<\tilde{h}_{x}(f)\tilde{h}^{*}_{y}(f')\right>=\frac{1}{2}\delta (f-f') \Gamma_{xy} S_{h}(f)\;, \quad \textrm{where}\; S_{h}(f)=\frac{A^{2}}{12\pi^{2}f_{0}^{3}}\left(\frac{f}{f_{0}}\right)^{-\gamma}\; .\end{equation}
Note that $\tilde{h}_{x}(f)$ here refers to the Fourier transform of the timing residuals in pulsar $x$, not the Fourier transform of the underlying tensor field. A similar calculation to Eq.\ (\ref{eq:whatisop}) shows that the optimal filter function is now given by
\begin{equation}\label{eq:opfilterstoch} \tilde{{\bf{Q}}}(f)=\frac{T S_{h}(f)}{S_{n}^{2}}{\bf{\Gamma}} \; , \quad \textrm{where}\; \left({\bf{\Gamma}}\right)_{xy}=\Gamma_{xy} .\end{equation}
As in Sec.\ \ref{sec:mono_freq}, the SNR of the statistic is given by the ratio of the expectation value in the presence of a signal to the rms value in the absence of a signal,
\begin{equation}\label{eq:snrback}  \rho^{2}= \sum_{x>y}\sum_{y}8T\int\textrm{d}f\;\frac{\Gamma_{xy}^{2}S_{h}^{2}(f)}{S^{2}_{n}(f)} \;.\end{equation}
Note that the continuous-wave SNR is different from the stochastic GW background SNR, i.e. $\varrho \neq \rho$. Since the pulsar term is being neglected the {\it overlap reduction function}, $\Gamma_{xy}$, is independent of frequency. For an isotropic background, the value of $\Gamma_{xy}$ depends only on the angle between the pulsar, $\cos(\theta_{xy})=\hat{p}_{x}\cdot\hat{p}_{y}$: this is the famous ``{\it Hellings and Downs curve}'' \citep{hellings-downs-1983}. As the pulsars in our PTA are randomly placed on the sky the overlap can be approximated as a constant, $\chi '$, equal to the rms value over the sky, $\Gamma_{xy}(\theta_{xy})=\chi ' =1/\left(4\sqrt{3}\right)$.

When searching for stochastic GW backgrounds of the form in Eq.\ (\ref{eq:OverlapReductionFunction}) all frequencies contribute to the signal. It no longer makes sense to ask what the sensitivity is as a function of frequency. Rather, the sensitivity in terms of $\gamma$ (or $\alpha=(3-\gamma)/2$) should be determined.
Substituting Eqs.\ (\ref{eq:OverlapReductionFunction}) and \ref{eq:opfilterstoch} into Eq.\ (\ref{eq:snrback}) gives
\begin{equation}\rho^{2}= \frac{1}{2}N_{p}\left(N_{p}-1\right)T\int\textrm{d}f\;\frac{\chi '^{2}A^{4}f^{4\alpha -6}}{4\sigma^{4}\delta t^{2}f_{0}^{4\alpha}}\; . \end{equation}
Setting $\rho=\rho_{\textrm{th}}$ gives an expression for $A$ in terms of $\alpha$, then for many values of $\alpha$ the curve $h_{c}(f)$ may be drawn using 
\begin{equation} h_{c}(f)= A\left(\frac{f}{f_{0}}\right)^{\alpha} \quad \textrm{where}\;\gamma=3-2\alpha\;. \end{equation}
This is plotted in the left panel of Fig.\ \ref{fig:stoch_comparison} using the same values for the PTA parameters as used in Fig.\ \ref{fig:mono_comparison}. These are the power-law integrated sensitivity curves of \citet{thraneromano2013}. 

\subsection{Bayesian detection}\label{sec:stoch_baye}
As in Sec.\ \ref{sec:mono_baye} the evidence for two competing hypotheses may be calculated; the signal hypothesis (${\cal{H}}_{h}$: the measured signal consists of a GW component, noise and the timing model) and the noise hypothesis (${\cal{H}}_{n}$: the measured signal consists of just noise and the timing model). 

As in section \ref{sec:mono_baye}, the noise is Gaussian, white and uncorrelated between pulsar, so has the same diagonal covariance matrix, ${\bf{\Sigma}}_{n}$, as before. From Eq.\ \ref{eq:OverlapReductionFunction} the Signal has covariance matrix
\begin{equation} {\bf{\Sigma}}_{h}= \left(\begin{matrix}{\bf{C}}&\Gamma_{12}{\bf{C}}&\ldots&\Gamma_{1N_{p}}{\bf{C}}\\
\Gamma_{21}{\bf{C}}&{\bf{C}}&\ldots&\Gamma_{2N_{p}}{\bf{C}}\\
\vdots&\vdots&\ddots&\vdots\\
\Gamma_{N_{p}1}{\bf{C}}&\Gamma_{N_{p}2}{\bf{C}}&\ldots&{\bf{C}}
\end{matrix}\right) \; , \end{equation}
\begin{equation} \textrm{where}\quad{\bf{C}}=\left(\begin{matrix}  
c(0)&c(2\pi\delta t)&\ldots&c(2\pi (T-\delta t))\\
c(2\pi\delta t)&c(0)& \ldots&c(2\pi(T-2\delta t))\\
\vdots&\vdots&\ddots&\vdots\\
c(2\pi (T-\delta t))&c(2\pi(T-2\delta t))&\ldots&c(0)\end{matrix}\right) \; , \end{equation}
which depends on the auto-correlation function of the timing residuals, given by
\begin{equation} c(\tau)=\int_{f_{\textrm{low}}}^{f_{\textrm{high}}}\textrm{d}f\; \cos\left(\tau f\right)S_{h}(f)^{-\gamma} \; . \end{equation}
In the signal hypothesis, the measured data is the sum of the noise and the signal. Since these are both drawn from a zero-mean Gaussian distributions the resulting distribution is also a zero mean and Gaussian with a covariance matrix given by ${\bf{\Sigma}}_{n}+{\bf{\Sigma}}_{h}$. Using the result in Eq.\ \ref{eq:On} the likelihood for the noise and signal hypotheses are given respectively by
\begin{eqnarray} \;\;{\cal{L}}_{n}&=\frac{\exp\left(-\frac{1}{2}{\bf{s}}^{\textrm{T}}{\bf{G}}\left({\bf{G}}^{\textrm{T}}{\bf{\Sigma}}_{n} {\bf{G}}\right)^{-1}{\bf{G}}^{\textrm{T}}{\bf{s}} \right)}{\sqrt{(2\pi)^{\xi}\textrm{det}\left({\bf{G}}^{\textrm{T}}{\bf{\Sigma}}_{n}{\bf{G}}\right)}}\; , \\
{\cal{L}}_{h}(\vec{\Psi})&=\frac{\exp\left(-\frac{1}{2}{\bf{s}}^{\textrm{T}}{\bf{G}}\left({\bf{G}}^{\textrm{T}}\left({\bf{\Sigma}}_{n} +{\bf{\Sigma}}_{h}\right) {\bf{G}}\right)^{-1}{\bf{G}}^{\textrm{T}}{\bf{s}} \right)}{\sqrt{(2\pi)^{\xi}\textrm{det}\left({\bf{G}}^{\textrm{T}}\left({\bf{\Sigma}}_{n} +{\bf{\Sigma}}_{h}\right){\bf{G}}\right)}}\; , \nonumber\end{eqnarray}
where $\xi=N_{p}(T/\delta t - 3)$. As in Sec.\ \ref{sec:mono_baye} the evidence for each hypothesis is calculated by integrating the prior-weighted likelihood over all the free parameters in the hypothesis. For ${\cal{H}}_{n}$ there are no free parameters and we simply have ${\cal{O}}_{n}={\cal{L}}_{n}$. For ${\cal{H}}_{h}$ we have the free parameters $\vec{\Psi}=(A,\alpha)$, adopting a delta function prior on both of these parameters gives ${\cal{O}}_{h}={\cal{L}}_{h}$. The Bayes factor is given by
\begin{eqnarray} {\cal{B}}= &\sqrt{\frac{\textrm{det}\left({\bf{G}}^{\textrm{T}}{\bf{\Sigma}}_{n}{\bf{G}}\right)}{\textrm{det}\left({\bf{G}}^{\textrm{T}}\left({\bf{\Sigma}}_{n}+{\bf{\Sigma}}_{h}\right){\bf{G}}\right)}}\times\\
&\exp\left(-\frac{1}{2}{\bf{s}}^{\textrm{T}}\left[{\bf{G}}\left({\bf{G}}^{\textrm{T}}\left({\bf{\Sigma}}_{n}+{\bf{\Sigma}}_{h}\right){\bf{G}}\right)^{-1}{\bf{G}}^{\textrm{T}}-{\bf{G}}\left({\bf{G}}^{\textrm{T}}{\bf{\Sigma}}_{n}{\bf{G}}\right)^{-1}{\bf{G}}^{\textrm{T}}\right]{\bf{s}}\right)\; .\nonumber\end{eqnarray}
Averaging the Bayes factor over many signal realisations gives the expectation value of the Bayes factor as $\overline{{\cal{B}}}$,
\begin{eqnarray} &P({\bf{s}})\textrm{d}{\bf{s}}=\frac{\exp\left(-\frac{1}{2}{\bf{s}}\left({\bf{\Sigma}}_{n}+{\bf{\Sigma}}_{h}\right)^{-1}{\bf{s}}\right)}{\sqrt{(2\pi)^{N_{p}T/\delta t}\textrm{det}\left({\bf{\Sigma}}_{n}+{\bf{\Sigma}}_{h}\right)}}\,\textrm{d}{\bf{s}}\, , \end{eqnarray}
\begin{equation}\label{eq:matrixBth} \overline{{\cal{B}}}=\frac{\sqrt{\textrm{det}\left({\bf{G}}^{\textrm{T}}{\bf{\Sigma}}_{n}{\bf{G}}\right)\textrm{det}\left(\left(2\left({\bf{G}}^{\textrm{T}}\left({\bf{\Sigma}}_{n}+{\bf{\Sigma}}_{h}\right){\bf{G}}\right)^{-1}-\left({\bf{G}}{\bf{\Sigma}}_{n}{\bf{G}}\right)^{-1}\right)^{-1}\right)}}{\textrm{det}\left({\bf{G}}^{\textrm{T}}\left({\bf{\Sigma}}_{n}+{\bf{\Sigma}}_{h}\right){\bf{G}}\right)} \, . \end{equation}
The matrix ${\bf{\Sigma}}_{h}$ depends of $A$ and $\gamma$, so by setting $\overline{{\cal{B}}}={\cal{B}}_{\textrm{th}}$ this expression may be solved to find $A$ in terms of $\gamma$. In the general case this is a function of large, dense, matrices and must be evaluated numerically.

\begin{figure}[h!]
 \centering
 \includegraphics[trim=0cm 0cm 0cm 0cm, width=0.9\textwidth]{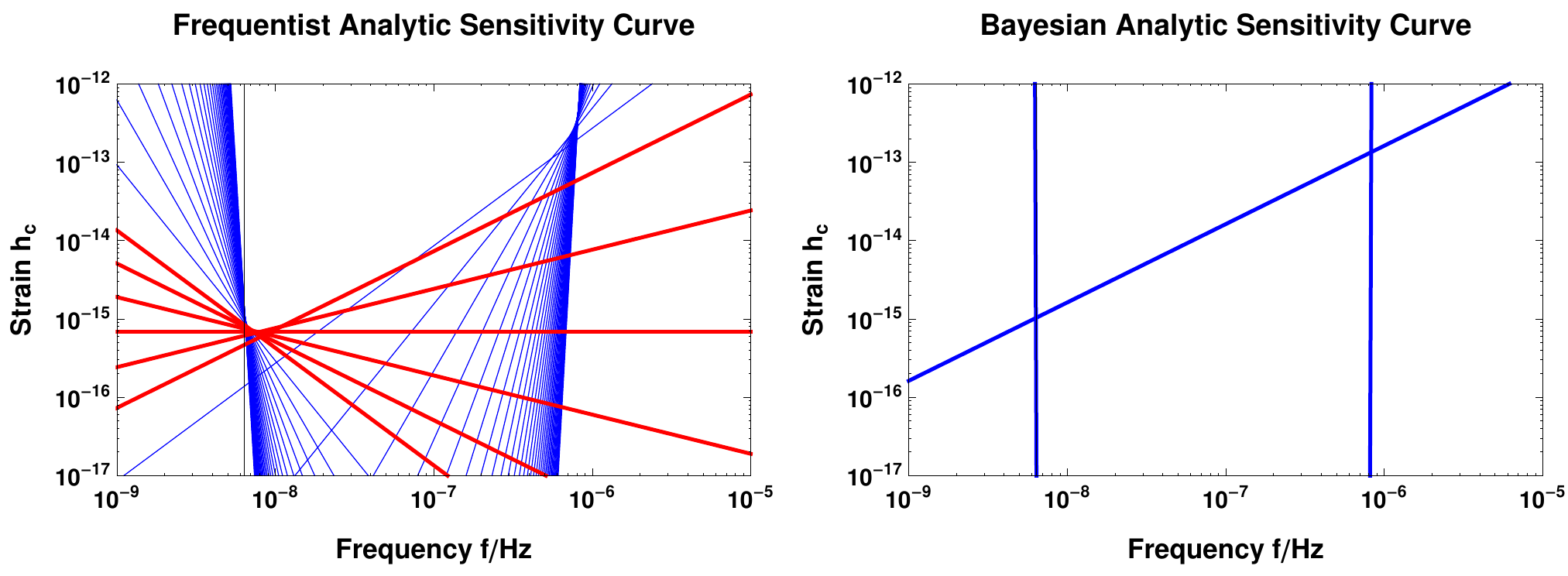}
 \caption{Power-law integrated sensitivity curves for a PTA's response to a stochastic GW background. The left panel shows the prediction of the frequentist formula in Sec.\ \ref{sec:stoch_freq}, and the right panel shows the prediction of the Bayesian formula in Sec.\ \ref{sec:stoch_baye}. The slopes shown in red are also plotted Fig.\ \ref{fig:stoch_surface}.}
 \label{fig:stoch_comparison}
\end{figure}

In the case of our very simple PTA where all of the pulsar are timed identically we can proceed a little further analytically. The matrix ${\bf{\Sigma}}_{h}$ is symmetric and the matrix ${\bf{\Sigma}}_{n}$ is isotropic, therefore we choose to evaluate Eq.\ \ref{eq:matrixBth} in the frame where both matrices are diagonal. As in section \ref{sec:stoch_freq} we approximate $\Gamma_{xy}=\chi'$ for all $x\neq y$, so ${\bf{\Sigma}}_{h}$ may be written in block diagonal form 
\begin{equation} {\bf{\Sigma}}_{h}^{\textrm{diag}}=\left(\begin{matrix}
\Lambda {\bf{C}}&0&\ldots&0\\
0&\lambda {\bf{C}}&\ldots&0\\
\vdots&\vdots&\ddots&\vdots\\
0&0&\ldots&\lambda {\bf{C}}\end{matrix}\right) \;\textrm{where,}\;\Lambda=1+(N_{p}-1)\chi ',\;\textrm{and}\;\lambda=1-\chi ' \; .\end{equation}
For the case of a white timing residual spectrum, $\gamma=0$, the matrix ${\bf{C}}$ is diagonal with identical entries $A^{2}a$, where $a=1/24\pi^{2}f_{0}^{3}\delta t$. In this case Eq.\ \ref{eq:matrixBth} simplifies to
\begin{eqnarray}&\label{eq:numericalBth} \overline{{\cal{B}}}_{\textrm{th}}=\frac{\sqrt{\sigma^{2\xi}\left(\frac{\sigma^{2}(\sigma^{2}+\Lambda A^{2} a)}{2\sigma^{2}-(\sigma^{2}+\Lambda A^{2} a)}\right)^{T/\delta t}\left(\frac{\sigma^{2}(\sigma^{2}+\lambda A^{2} a)}{2\sigma^{2}-(\sigma^{2}+\lambda A^{2} a)}\right)^{\xi-T/\delta t}}}{\left(\sigma^{2}+\Lambda A^{2}a\right)^{T/\delta t}\left(\sigma^{2}+\lambda A^{2}a\right)^{\xi-T/\delta t}}\, .\end{eqnarray}
Eq.\ \ref{eq:numericalBth} may be solved for $A$ using a simple root finding algorithm. From the frequentist analysis in Sec.\ \ref{sec:stoch_freq} and the left panel of Fig.\ \ref{fig:stoch_comparison} it is expected that the power law integrated sensitivity curve  will be vertical at frquencies of $1/T$ and $1/\delta t$, and joined by a curve $h_{c}\propto f$ (i.e $\gamma=1$). Therefore, in order to uniquely determine the sensitivity curve all that must be determined is the amplitude $A(\gamma =1)$. The numerically found root of Eq.\ \ref{eq:numericalBth} gives us $A(\gamma = 0)$. Approximating $A(\gamma=1)\approx A(\gamma=0)$ gives the curve shown in the right panel of Fig.\ \ref{fig:stoch_comparison}. The justification for this crude approximation comes \emph{a posteriori} from the good agreement which can be seen in Figs.\ \ref{fig:stoch_comparison} and \ref{fig:stoch_surface} and the relatively week dependance of $A$ on $\gamma$ for small values of the slope which can be seen in the left panel of Fig.\ \ref{fig:stoch_comparison}.

\subsection{Numerical calculations}\label{sec:stoch_num}
The shape of the strain sensitivity curve describing the PTA response to an isotropic, stationary, Gaussian stochastic GW background is now reconstructed with numerical simulations. Our canonical PTA is as in Sec.\ \ref{sec:mono_num}. We inject isotropic GW background signals using the \textbf{GWbkgrd} plugin \citep{tempo2-3} for \textsc{Tempo2}, which simulates a large number ($\sim \mathcal{O}(10^4)$) of GW sources between frequencies much less than $\sim 1/T$ and much greater than Nyquist. Plus and cross GW polarisation amplitudes are drawn according to the user-specified spectral shape, corresponding to a choice of $A$ and $\alpha$, and the resulting TOA deviations induced by each oscillator are summed to produce a residual time-series. The timing-files are processed with \textsc{Tempo2} as in the single-source simulations, and the likelihood is again marginalised over uniform-prior timing-model parameters.

\begin{figure}[h!]
 \centering
 \includegraphics[trim=0cm 0cm 0cm 0cm, width=0.9\textwidth]{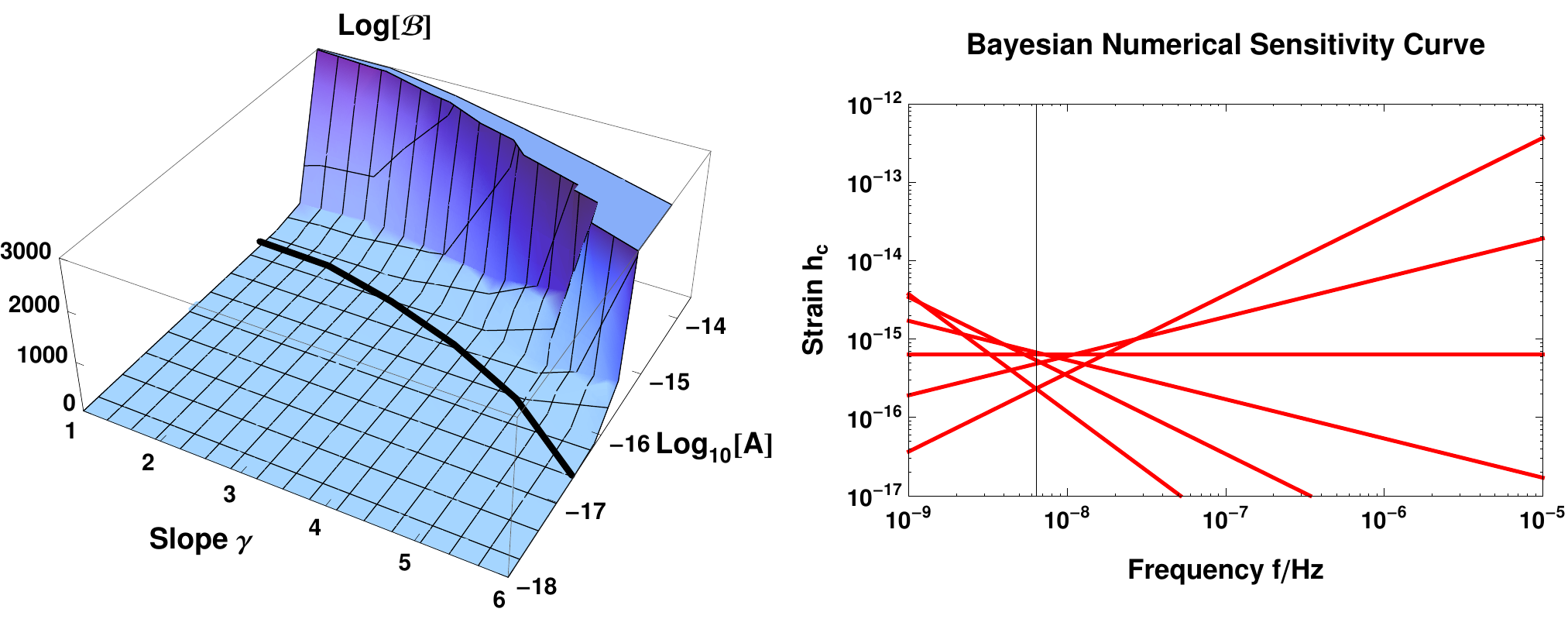}
 \caption{The left panel shows a plot of $\log ({\cal{B}})$ against amplitude and slope, the black line indicates the detection threshold. The right panel shows the corresponding power-law integrated sensitivity curve discussed in Sec.\ \ref{sec:stoch_num}.}
 \label{fig:stoch_surface}
\end{figure}

For $\alpha\in\left\{-1.5,-1.0,-0.5,0,0.5,1.0\right\}$  stochastic background signals of varying SNR were injected by scaling $A$. We note that the expected spectral-slope for a background composed of purely GW-driven inspiraling SMBH binaries is $\alpha \sim -2/3$ \citep[e.g.,][]{begelman1980}. Our likelihood model is the composite time-frequency approach of \citet{lentati-timefrequency-2013} which accelerates the computationally expensive linear algebra operations necessary within the usual time-domain stochastic background search. Parameter estimation and evidence recovery with \textsc{MultiNest} gives us a collection of signal/noise evidences for each combination of injected $(A,\alpha)$. As before, at each $\alpha$ the recovered Bayes factors were interpolated to give the corresponding value of $A$ required to exceed our pre-determined detection threshold. The results are shown in Fig.\ \ref{fig:stoch_surface}. The results show comfortable qualitative agreement with a simple frequentist/Bayesian analytic techniques, although reconstruction is limited by the range of injected values of $\alpha$ which was necessary for numerical stability.

The r.m.s value of the overlap between pulsars in the mock dataset used is $0.25$. This disagrees with the expected value $\chi '=1/(4\sqrt{3})\approx 0.144$ calculated in Sec.\ \ref{sec:stoch_freq} by a factor of $\approx 2$; this simply represents the failure of a particular realisation of 36 pulsars on the sky to give the mean of a distribution of a large number of point drawn randomly on a sphere.

\section{Discussion}\label{sec:discussion}
Here the sensitivity curve of a canonical PTA roughly equivalent to mock dataset \textsc{Open1} in the recent IPTA data challenge has been calculated for both a monochromatic wave and a power-law stochastic background. These calculations have been performed in both Bayesian and frequentist frameworks and using both analytic and numerical techniques. The results show excellent qualitative agreement and good quantitative agreement up to a factor of a few, which is as much as could be expected considering the differing assumptions necessary in each calculation. Additionally we have presented several simple analytic formulae for both monochromatic and stochastic background sensitivity curves in both the Bayesian and frequentist pictures. Along the way to deriving these sensitivity curves simple analytic formulae for the frequentist signal-to-noise-ratio and the Bayesian evidence have also been derived.

The different sensitivity curves for the monochromatic wave and stochastic background illustrate the fact that sensitivity curve of a PTA depends both on the properties of the source and the properties of the measured pulse TOA dataset. This difference is not specific to PTAs, it is also present in ground and space-based GW detectors. However the differences are particularly pronounced in the case of PTAs because they are most sensitive to frequencies $\sim 1/T$, where $T$ is the total baseline observation time, and hence there are only a few complete wave cycles in the data. This is in contrast to, say, LIGO, which has a peak sensitivity of $\sim 100\,\textrm{Hz}$, so a year's data contains $\sim 10^{9}$ cycles.

\section*{Acknowledgements}
This work was performed using the Darwin Supercomputer of the University of Cambridge High Performance Computing Service (http://www.hpc.cam.ac.uk/), provided by Dell Inc. using Strategic Research Infrastructure Funding from the Higher Education Funding Council for England and funding from the Science and Technology Facilities Council. Both CM and ST are supported by the STFC. JG's work is supported by the Royal Society.

\appendix
\section{Bayesian monochromatic sensitivity}\label{app}
Rearranging Eq.\ \ref{eq:final} gives the Bayesian expression for the sensitivity of a PTA to a monochromatic source.
\begin{eqnarray}
h_{c}&=4 \chi  \sqrt{\log ({\cal{B}})} \left(-\frac{90 N_{p} \cos ^2(\phi )}{\pi ^6 \delta t f^8 \sigma ^2 T^5}-\frac{90 N_{p} \cos ^2(2 \pi 
   f T+\phi )}{\pi ^6 \delta t f^8 \sigma ^2 T^5}+\frac{180 N_{p} \cos (\phi ) \cos (2 \pi  f T+\phi )}{\pi ^6 \delta t f^8 \sigma ^2
   T^5}\right.\nonumber \\
&\left. +\frac{180 N_{p} \sin (\phi ) \cos (\phi )}{\pi ^5 \delta t f^7 \sigma ^2 T^4}+\frac{180 N_{p} \cos (\phi ) \sin (2 \pi  f T+\phi )}{\pi ^5
   \delta t f^7 \sigma ^2 T^4}-\frac{180 N_{p} \sin (\phi ) \cos (2 \pi  f T+\phi )}{\pi ^5 \delta t f^7 \sigma ^2 T^4}\right.\nonumber\\
&\left.-\frac{180 N_{p}
  \sin (2 \pi  f T+\phi ) \cos (2 \pi  f T+\phi )}{\pi ^5 \delta t f^7 \sigma ^2 T^4}-\frac{96 N_{p} \sin ^2(\phi )}{\pi ^4 \delta t f^6
  \sigma ^2 T^3}-\frac{96 N_{p} \sin ^2(2 \pi  f T+\phi )}{\pi ^4 \delta t f^6 \sigma ^2 T^3}\right.\nonumber\\
&\left.-\frac{168 N_{p} \sin (\phi ) \sin (2 \pi  f T+\phi
   )}{\pi ^4 \delta t f^6 \sigma ^2 T^3}+\frac{120 N_{p} \cos (\phi ) \sin (\pi  f T) \sin (\pi  f T+\phi )}{\pi ^4 \delta t f^6 \sigma ^2
   T^3}\right.\nonumber\\
&\left.-\frac{120 N_{p} \sin (\pi  f T) \sin (\pi  f T+\phi ) \cos (2 \pi  f T+\phi )}{\pi ^4 \delta t f^6 \sigma ^2 T^3}-\frac{144 N_{p} \sin
   (\phi ) \sin (\pi  f T) \sin (\pi  f T+\phi )}{\pi ^3 \delta t f^5 \sigma ^2 T^2}\right.\nonumber\\
&\left.-\frac{96 N_{p} \sin (\pi  f T) \sin (\pi  f T+\phi ) \sin (2 \pi 
   f T+\phi )}{\pi ^3 \delta t f^5 \sigma ^2 T^2}-\frac{24 N_{p} \sin (\phi ) \cos (2 \pi  f T+\phi )}{\pi ^3 \delta t f^5 \sigma ^2
   T^2}\right.\nonumber\\
&\left.+\frac{24 N_{p} \sin (2 \pi  f T+\phi ) \cos (2 \pi  f T+\phi )}{\pi ^3 \delta t f^5 \sigma ^2 T^2}-\frac{72 N_{p} \sin ^2(\pi  f T) \sin
   ^2(\pi  f T+\phi )}{\pi ^2 \delta t f^4 \sigma ^2 T}\right.\nonumber\\
&\left.-\frac{24 N_{p} \cos ^2(2 \pi  f T+\phi )}{\pi ^2 \delta t f^4 \sigma ^2 T}-\frac{48
   N_{p} \sin (\pi  f T) \sin (\pi  f T+\phi ) \cos (2 \pi  f T+\phi )}{\pi ^2 \delta t f^4 \sigma ^2 T}\right.\nonumber\\
&\left. +\frac{N_{p} \sin (2 \phi )}{\pi 
   \delta t f^3 \sigma ^2} -\frac{N_{p} \sin (2 (2 \pi  f T+\phi ))}{\pi  \delta t f^3 \sigma ^2}+\frac{4 N_{p} T}{\delta t f^2
   \sigma ^2}\right)^{-1/2}
\end{eqnarray}

\bibliographystyle{apj}
\bibliography{bibliography,apj-jour}    %% includes the journal abbrevs
%\bibliography{bibliography}

\end{document}